\documentclass[cits]{PoS}

\usepackage{graphicx}
\usepackage{times}
\usepackage{amssymb}
\usepackage{url}
\usepackage[authoryear]{natbib}
\bibpunct{(}{)}{;}{a}{}{,}

\title{Probing the neutron star interior and the Equation of State of cold dense matter with the SKA}

\ShortTitle{The neutron star EOS}

\author{\speaker{Anna Watts}$^1$, Renxin Xu$^2$, Crist\'obal Espinoza$^3$, Nils Andersson$^4$, John Antoniadis$^{5}$, Danai Antonopoulou$^1$, Sarah Buchner$^7$, Shi Dai$^8$, Paul Demorest$^9$, Paulo Freire$^6$, Jason Hessels$^{1,10}$, J\'er\^ome Margueron$^{11}$, Micaela Oertel$^{12}$, Alessandro Patruno$^{13,10}$, Andrea Possenti$^{14}$, Scott Ransom$^{9}$, Ingrid Stairs$^{15}$, Ben Stappers$^{16}$\\  

$^1$University of Amsterdam;
$^2$KIAA Peking University; 
$^3$PUC, Chile;
$^4$University of Southampton;
$^5$University of Toronto
$^6$ Max Planck Institute for Radio Astronomy;
$^7$ HRAO, University of Witwatersrand;
$^8$ ATNF;
$^9$ NRAO Charlottesville;
$^{10}$ ASTRON;
$^{11}$ IPNL, CNRS/IN2P3, Universit\'e Lyon;
$^{12}$ LUTH, CNRS, Observatoire de Paris, Universit\'e Paris Diderot;
$^{13}$ Leiden University;
$^{14}$ INAF-Osservatorio Astronomico di Cagliari;
$^{15}$ University of British Columbia;
$^{16}$ University of Manchester.

E-mail: \email{A.L.Watts@uva.nl}}

\abstract{With an average density higher than the nuclear density, neutron stars (NS) provide a unique test-ground for nuclear physics, quantum chromodynamics (QCD), and nuclear superfluidity.  Determination of the fundamental interactions that govern matter under such extreme conditions is one of the major unsolved problems of modern physics, and -- since it is impossible to replicate these conditions on Earth -- a major scientific motivation for SKA.  The most stringent observational constraints come from measurements of NS bulk properties: each model for the microscopic behaviour of matter predicts a specific density-pressure relation (its `Equation of state', EOS). This generates a unique mass-radius relation which predicts a characteristic radius for a large range of masses and a maximum mass above which NS collapse to black holes. It also uniquely predicts other bulk quantities, like maximum spin frequency and moment of inertia.   The SKA, in Phase 1 and particularly in Phase 2 will, thanks to the exquisite timing precision enabled by its raw sensitivity, and surveys that dramatically increase the number of sources:  1) Provide many more precise NS mass measurements (high mass NS measurements are particularly important for ruling out EOS models); 2) Allow the measurement of the NS moment of inertia in highly relativistic binaries such as the Double Pulsar; 3) Greatly increase the number of fast-spinning NS,  with the potential discovery of spin frequencies above those allowed by some EOS models; 4) Improve our knowledge of new classes of binary pulsars such as black widows and redbacks (which may be massive as a class) through sensitive broad-band radio observations; and 5) Improve our understanding of dense matter superfluidity and the state of matter in the interior through the study of rotational glitches, provided that an ad-hoc campaign is developed.

}

\FullConference{
Advancing Astrophysics with the Square Kilometer Array\\
June 8-13, 2014\\
Giardini Naxos, Italy}

\begin{document}

\section{Introduction}
\label{intro}

\begin{figure}
\centering
\includegraphics[width=14cm]{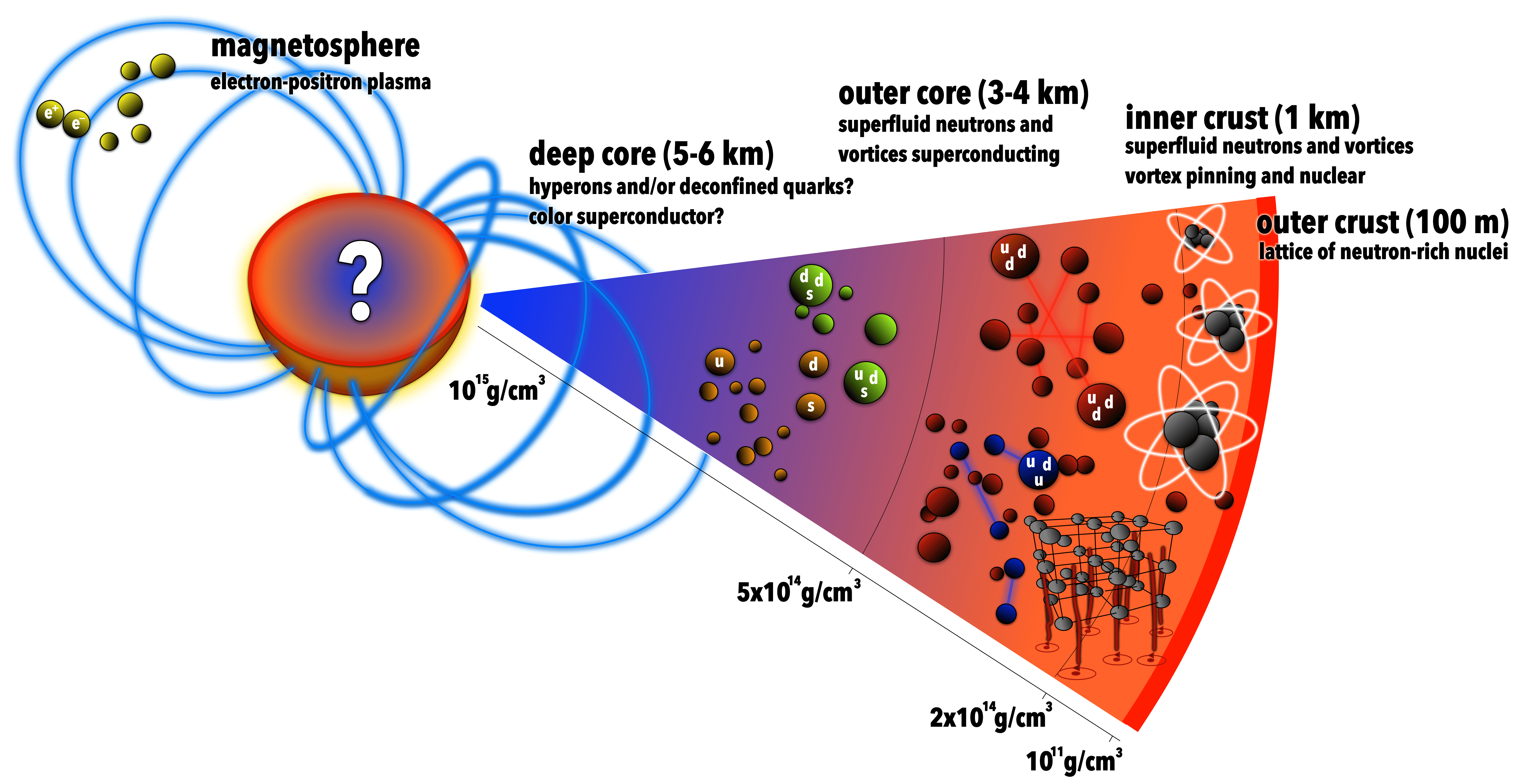}
\caption{Schematic structure of a NS. The outer layer is a solid ionic crust supported by electron degeneracy pressure. Neutrons begin to leak out of nuclei at
densities $\sim4\times10^{11}$ g/cm$^3$ (the neutron drip line, which separates inner and outer crust), where neutron degeneracy also starts to play a role. At densities $\sim 2\times 10^{14}$  g/cm$^3$, the crust-core boundary, nuclei dissolve completely. In the core, densities may reach up to ten times the
nuclear saturation density $\rho_\mathrm{sat} = 2.8\times10^{14}$ g/cm$^3$ (the density in normal atomic nuclei). }
\label{schema}
\end{figure}

Neutron stars (NS) are the densest objects in the Universe.  Composition evolves from ions embedded in a sea of degenerate electrons (and eventually neutrons) in the solid crust, through the crust-core transition where nuclei dissolve to form neutron-rich nucleonic matter, to the supranuclear densities of the core, where theory suggests the presence of exotic non-nucleonic matter (Figure \ref{schema}).  The
nature of matter in such extreme conditions is one of the great unsolved problems in modern science, making NS unparalleled laboratories for nuclear physics and quantum chromodynamics (QCD). NS also host super-strong internal magnetic fields, and share the rich phenomenology of low-temperature physics.   Although born in the hot furnace of a supernova collapse, NS cool rapidly to temperatures far below the relevant Fermi temperature. When matter cools there are two options:  it can freeze to form a solid, or become superfluid (or superconducting if there are free charge carriers involved).  NS are expected to exhibit both these phases.  The outer kilometre or so of the star forms a solid crustal lattice of increasingly neutron-rich nuclei, while the star's core hosts a neutron superfluid coupled to a proton superconductor and a free gas of electrons (as required to make the conglomerate charge neutral).  The free neutrons permeating the crustal lattice in the inner crust are also expected to be superfluid.   

The most fundamental macroscopic diagnostic of dense matter interactions is the pressure-density-temperature
relation of bulk matter, the equation of state (EOS). The EOS can be used to infer key aspects of
the microphysics, such as the nature of the three-nucleon interaction or the presence of free quarks at high
densities. Determining the EOS of supranuclear density matter is therefore of major importance to
fundamental physics. However it is also critical to astrophysics, to understand not only NS but also NS/NS and NS/Black Hole (BH) mergers, prime sources of gravitational waves and the likely engines of short gamma-ray bursts \citep{Bauswein2012,Lackey2012}. The EOS affects merger dynamics, BH formation timescales, the precise gravitational wave and neutrino signals, any associated mass loss and
r-process nucleosynthesis, and the attendant gamma-ray bursts and optical flashes \citep{Metzger2010,Hotokezaka2011}. The cold EOS probed with NS, and the way it joins up with the hot EOS that determines explosion conditions \citep{Janka2007}, are also vital to understanding the late
stages of core collapse supernovae, including their gravitational wave signal.

\subsection{Unknowns in dense matter physics}
\label{unknowndm}

The properties of NS, like those of atomic nuclei, depend crucially on the
interactions between protons and neutrons (nucleons), which are a starting point
for ab-initio calculations of the nuclear many-body problem.  While
two-body interactions are well constrained by experiment, three- and more-body
forces are a frontier in nuclear physics. At low energies, effective field
theories based on QCD symmetries provide a systematic expansion of nuclear
forces, which predict two- and more-nucleon interactions. There are also
complementary efforts using lattice approaches to nuclear forces to provide few-body
nucleon-nucleon and more generally baryon-baryon interactions, but this
approach is still affected by large uncertainties.  Some of the predictions of these models can be tested against current nuclear data; where data are not yet available, predictions are 
based on the consistency of the approach. The appearance of shell closure in
neutron-rich isotopes, as well as the position of the neutron drip-line is
found to be very sensitive to the three-body forces.  Exotic neutron-rich
nuclei, the focus of present and upcoming laboratory experiments, also provide interesting constraints on effective interactions for many-body systems. 

\begin{figure}
\centering
\includegraphics[width=15cm]{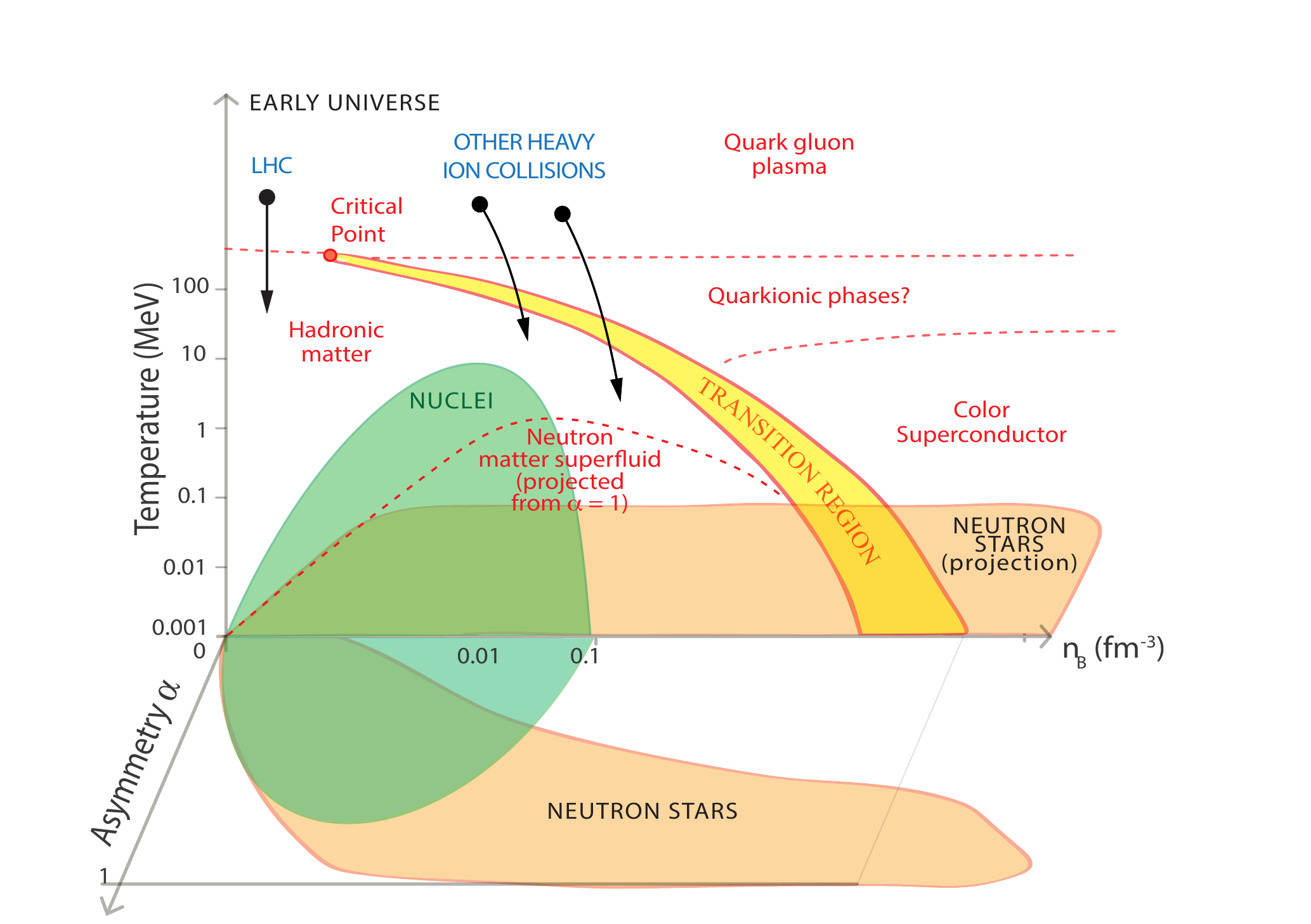}
\caption{The parameter space and the states of matter accessed by NS, as compared to laboratory experiments on Earth. The figure shows temperature $T$ against baryon number density $n_B$ against asymmetry $\alpha = 1 - 2 Y_q$, where $Y_q$ is the hadronic charge fraction (in general number of protons/number of baryons).  $\alpha = 0$ for matter with equal numbers of neutrons and protons, and $\alpha = 1$ for pure neutron matter.  Green region: Isolated nuclei up to $\alpha \approx 0.3$, above this one would find a mix of nuclei and light particles.}
\label{trho}
\end{figure}

NS observations, by contrast, challenge many-nucleon interactions at
extremes of density, neutron richness, and baryonic content.  Astrophysical
inputs are essential to make progress (Figure \ref{trho}), although extracting information from the
data is harder than for lab experiments.  Current approximate models all have major
uncertainties, both at high density and for neutron-rich matter \citep{Wiringa1995,Pieper2001,Stone2007,Hebeler2010,Steiner2012}.  Although experimental information on matter 
near the nuclear saturation density is plentiful, there are only a few
experimental constraints at high densities and/or for neutron-rich matter \citep{Tsang2009,Kortelainen2010,RocaMaza2011,Piekarewicz2012}. NS thus provide a unique environment to test models.

At the very highest densities reached in NS cores, we also
expect transitions to non-nucleonic states of matter \citep{Glendenning2000}. Some of
the most exciting possibilities involve strange quarks: unlike heavy
ion collision experiments, which always produce very short-lived and
hot, dense states, the stable gravitationally confined environment of a
NS permits slow-acting weak interactions that can form
matter with a high net strangeness. Possibilities include
the formation of hyperons (strange baryons), free quarks (forming a
hybrid star), or color-superconducting states \citep{Alford2008}. It is even possible
that the entire star converts into a lower energy self-bound state
consisting of up, down and strange quarks, known as a strange quark
star \citep{Witten1984,Weber2005}, possibly in a solid state \citep{Xu2003}. Other states that
have been hypothesized include Bose-Einstein condensates of mesons
(pions or kaons, the latter containing a strange quark) and $\Delta$
baryons (resonant states). The densities at which such phases would
appear, and the degree to which they might co-exist with other phases,
are highly uncertain. 

In principle, all these phases are governed by the strong
force. First principle calculations of the interactions for many-body
QCD systems are however presently unfeasible due to the fermion sign
problem. Because fermion wave functions change sign upon exchange of
particles, computing the expected energy of a system involves the
addition of a large number of terms with alternating signs. This is
especially problematic at high density, and it is at present
impossible to obtain direct predictions for strongly interacting quark
matter. Instead one has to resort to phenomenological models. This is
particularly true for hadronic many-body systems, since a
single hadron is already a complex many-body QCD system. Here, the
different models are all based on hadronic interactions. For the
non-nucleonic degrees of freedom such as hyperons, experimental information is scarce and uncertainties even
larger than for purely nuclear systems. Lattice approaches and effective field
theory expansion could help in the future to improve the starting point of
ab-initio hadronic many-body calculations with these degrees of
freedom. Planned laboratory experiments on hypernuclei (nuclei containing
at least one hyperon) will also give additional constraints. 

Figure \ref{trho} compares the parameter space that can be accessed within the
laboratory to that explored with NS. While the two are complementary, it is obvious that NS provide a
truly unique exploration space. In addition to the high densities, the
gravitational confinement in NS renders them sufficiently
long-lived such that weak beta equilibrium is achieved and as far as the EOS
is concerned, NS can be considered as cold.  Thus, neither the neutron rich
ground state of beta equilibrated ultra-dense nuclear matter nor the exotic
non-nucleonic states of matter described in the previous paragraph can be
reached in the laboratory.

\subsection{Unknowns in dense matter superfluidity}
\label{unknownsf}
 
The presence of superfluid components has significant effects on the star's long term evolution and dynamics. Superfluidity suppresses the nuclear reactions that lead to cooling, but also leads to new neutrino emission channels. The suggestion of rapid cooling of the young NS in the Cassiopeia A supernova remnant \citep{Page2011,Shternin2011} has stimulated a detailed discussion of the impact on superfluid parameters. If confirmed, such a fast cooling epoch would constrain the critical temperature at which the neutrons in the core become superfluid. At the same time, the effect requires the onset of proton superconductivity to have taken place much earlier.  In pulsar timing, superfluidity is thought to be associated with the observed restlessness, ranging from the enigmatic glitches seen in young pulsars to the timing noise that appears to be generic in all systems. A superfluid rotates by forming an array of quantised vortices (Figure \ref{schema}). The motion of these tiny tornadoes determines the star's spindown behaviour, e.g. whether it proceeds smoothly or in an erratic fashion.  

Key issues involve the friction experienced by the vortices and whether or not  they are pinned to some other component in the system. Current models suggest that the most important source of friction is electron scattering off of magnetic fields associated with the vortices \citep{Alpar1984, Mendell1991, asc2006}. The outcome depends to a large extent on the so-called entrainment. This is a non-dissipative effect that determines the dynamical effective mass of the particles involved, effectively encoding how mobile the superfluid component is. In the core of the star, the entrainment arises due to the strong interaction \citep{Alpar1984,Borumand1996}, while in the inner crust it is due to Bragg scattering off of the nuclear lattice \citep{Chamel2006}. The interaction between magnetised vortices and the much more plentiful fluxtubes associated with the proton superconductor  (e.g. the star's bulk magnetic field) may also be dissipative \citep{Link2003}. In some regimes it may even cause pinning, which would link the star's spin evolution to changes in the magnetic field. In the star's crust, vortices are expect to pin to (or between) the lattice nuclei. As outlined below, precision timing provides a unique probe of the parameters associated with these phenomena, going far beyond the current constraint on the critical temperature from cooling data and the ballpark superfluid moment of inertia from glitch observations.

In the absence of friction and pinning, the neutron superfluid would be more or less decoupled from the charged components in the star. This would have a major effect on the moment of inertia associated with the spindown, and could in fact lead to a complicated evolution of the braking index as the star cools and the superfluid region grows \citep{Ho2012}. On the other hand, vortex pinning is a key requirement for the ``cartoon-level'' explanation for glitches, where a pinned component builds up an angular momentum reservoir as the system evolves. The stored angular momentum is then released in some large scale unpinning event and the pulsar is seen to spin up. The relaxation that follows should be determined by vortex friction \citep{Sidery2010} and the mechanism that repins the vortices to return the system to the point where it may glitch again. This is a very complicated story, and there has been precious little progress in our theoretical understanding of the glitch phenomenon since the first observations at the end of the 1960s. However, a well-designed SKA campaign could change the situation dramatically.  At the same time a concerted effort is, obviously, required on the theory side. Finally, superfluidity is also expected to have  impact on possible free precession. Significant vortex pinning would, in fact, prevent slow precession \citep{Jones2001,Link2003}. Instead, one would expect the precession period to be similar to the rotation period. Higher quality data may allow us to constrain --- or even rule out --- this possibility.

\section{Methodology}
\label{methodology}

\subsection{Connecting observables to physics using pulsar timing}
\label{pulsartiming}

\begin{figure}
\includegraphics[width=15cm]{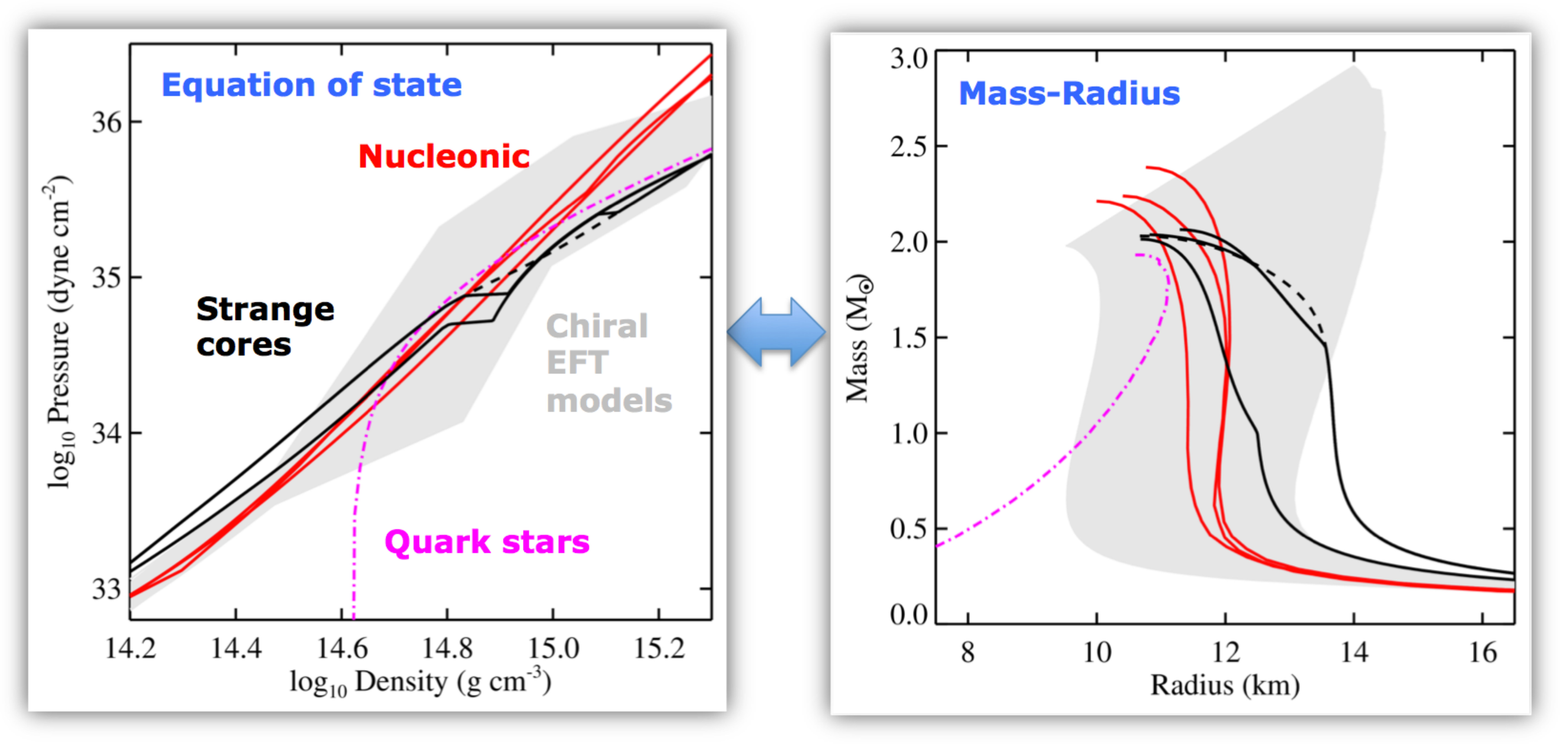}
\caption{The link between uncertainty in fundamental nuclear physics, the EOS, and the NS M-R relation.  The figure shows the pressure density relation (left) and the corresponding mass-radius relation (right) for different EOS models.  Grey band: range of nucleonic EOS based on chiral effective field theory \citep{Hebeler2013}. Black solid:  Hybrid EOS from \cite{Zdunik2013}. Black dashed:  Nucleon + hyperon core EOS from \cite{Bednarek2012}. The other EOS are from \cite{Lattimer2001}: Red (nucleonic); Magenta (quark stars). All models shown are compatible with the existence of 2 $M_\odot$ NS: discovery of a higher mass star would rule out the softer EOS with the lowest maximum masses.}
\label{eosmr}
\end{figure}

The relativistic equations of stellar structure\footnote{The Tolman-Oppenheimer-Volkoff equations (the equation of mass continuity, force balance, and the EOS P = P($\rho$) where $\rho$ is the total density and P the pressure), modified to take into account rotation.  The EOS should also include temperature: but for NS, we are so far below Fermi temperature that this can be neglected in computing bulk structure.}  relate the EOS of dense matter, which depends on the microphysics of composition and the strong force, to macroscopic observables including the mass M and radius R of the NS.  Figure \ref{eosmr} illustrates the relationship between the EOS and the M-R relation for different models:  there is a unique map between the two \citep{Lindblom1992}. Constraints on or measurements of M and/or R therefore limit EOS models, and hence the interior composition of NS.    Ideally, we would like to use the SKA to measure M and R for many NS  in order to trace
out the EOS in the M-R diagram.  Unfortunately, direct measurements of R in the radio are likely impossible, and even indirect measurements through estimates of the moment of inertia are, as discussed later in this chapter, incredibly difficult.  However measuring {\em masses} is something that
radio observations of pulsars, via high-precision timing measurements,
can do exceedingly well, and the SKA will be a fantastic tool for this
purpose.

Pulsar timing is the process of unambiguously accounting for every
rotation of a NS over extended time periods (as long as
decades, currently), and it provides astonishingly precise
measurements of the pulsar's spin, astrometric, and (if appropriate)
orbital parameters, including how each of these changes with time
\citep{lorimer08}.  For ``recycled'' pulsars, most of which are
in binary systems, the five Keplerian orbital parameters (orbital
period $P_b$, projected semi-major axis $x \equiv a_1 \sin i / c$
(where $i$ is the orbital inclination and $c$ is the speed of light),
eccentricity $e$, time of periastron passage $T_o$, and the argument
of periapsis $\omega$) are measured with typically 4$-$12 significant
figures, making them essentially perfectly known.

For some binary pulsars, one or more of the 
``post-Keplerian'' (PK) timing parameters may be measurable \citep{Shao14}.  These
observables (the precession of the argument of periapsis $\dot
\omega$, the orbital period decay $\dot P_b$, the combination of time
dilation and gravitational redshift $\gamma$, and the Shapiro delay
``range'' $r$ and ``shape'' $s$) can be expressed in the framework of
general relativity (GR) as functions of only physical constants (e.g.
solar mass in time units $T_{\odot} \equiv GM_{\odot}/c^3 \sim
5\,\mu$s), the extremely well-measured Keplerian parameters, and the
masses of the pulsar $m_p$ and the companion star $m_c$.  Since $m_p$
and $m_c$ are the only two unknowns, if two PK observables
are measured, the two equations can be used to solve for $m_p$ and
$m_c$.  Tests of GR come from measurements of additional PK
parameters.

Unfortunately, only a small fraction of binary pulsars have timing
precisions, orbital parameters, or companions which allow the
measurement of PK parameters.  To measure $\dot \omega$, for instance,
which by itself provides the total system mass $m_p + m_c$ and is
exactly the same effect as Mercury's precession of perihelion, we need
pulsars in eccentric enough orbits so that we can trace
$\omega$ precisely.  Most millisecond pulsars (MSPs) in the Galactic disk have
extremely circular orbits and so even though their orbital parameters
are measured to extraordinary precision, $\dot \omega$ cannot be.
However, globular clusters contain many eccentric MSPs \citep{ransom08,Hessels14}
and there is evidence for a population of them in the Galactic disk as
well \citep{champion+08,barr13}, providing opportunities for new
discoveries of eccentric systems by the SKA.

Detection of orbital period decay $\dot P_b$, due to the emission of
gravitational radiation, demands compact systems such as the original
Hulse-Taylor binary \citep{wnt10}.  These double NS (DNS) systems
\citep{Kramer2006,fst14} tend to be in highly eccentric orbits, which allows measurement of $\dot \omega$, and after enough
precession has occured, $\gamma$ (the requirement for measurable
precession makes $\gamma$ effectively impossible to measure in systems
with near-circular orbits, even if they are compact and relativistic).
Several compact NS---white dwarf (WD) systems have
been discovered recently which allow measurement of $\dot P_b$
\citep{bbv08,fwe+12,Antoniadis2013}, implying that the SKA should find
many more such systems.

The final PK parameters of interest are those making up the Shapiro
delay, which is a relativistic propagation delay, typically of the
order of $\mu$s, caused as the pulses travel through the gravitational
potential of the companion star.  The amplitude of this effect depends sensitively on the inclination angle and requires nearly edge on viewing and high timing precision to become measurable.
With huge improvements in timing precision over
the last decade, as well as the discovery of many new MSPs, the number
of Shapiro delay measurements has increased substantially, and that
number should increase tremendously with the SKA.  The first
high-precision 2$M_\odot$ NS was measured with this
technique \citep{Demorest2010}, and that single number has provided
exceptionally strong constraints on the EOS \citep{Hebeler2013}.

For non-binary slowly-spinning pulsars it is impossible to measure orbital
or PK parameters, yet rapid-cadence (daily or every few days)
timing observations can detect the timing ``glitches'' which young
pulsars often exhibit (Section \ref{sotaglitch}).  These virtually instantaneous increases in the
pulsar's spin rate and (sometimes) spin-down rate, and the subsequent relaxation process,
probe the superfluid interiors of NS.  

\subsection{Importance of large surveys to sample populations}
\label{surveys}

Surveys continue to be a major driver for pulsar science and its
applications.  By increasing the known population, we increase
the number of outlier systems whose exceptional properties make them
powerful physics laboratories, e.g.\citep{Kramer2006,Ransom2014}.
Consider the fact that there are 2300
known radio pulsars \citep{Manchester2005} out of an estimated total
Galactic population of a few tens of thousand sources beaming towards
the Earth \citep{Keane14}.  Of these known sources, fewer than 10\%
are in binary (or trinary) systems, a pre-requisite for
precision mass measurement through either Shapiro delay or a
combination of PK orbital parameters and constraints from
spectroscopic modeling of the optical companion star.  Unfortunately,
but unsurprisingly, only a handful of systems have extreme properties
that push the envelope in terms of constraining the EOS.  There are
only 10 radio pulsars with rotational periods less than 2ms, and only
2 NS with precision mass measurements of $M_{\rm NS}
\gtrsim 2$ M$_\odot$ \citep{Demorest2010,Antoniadis2013}.  Pushing to
faster spin rates, which can constrain the maximum NS
radius, and higher masses will require both a significant increase in
the known population (a doubling or tripling) as well as careful
follow-up and precision timing of the most promising discoveries.

In addition to strong constraints on the EOS from maximum masses, a much larger sample will also provide
precision mass measurements across the known distribution $M_{\rm NS}
\sim (1.2 - 2) M_\odot$ (lower mass limit from \cite{Janssen08,Ferdman14}).  Mapping the NS mass distribution \citep{Tauris14}
is important for understanding the {\it astro}physics of NS formation in
core-collapse supernovae (EOS dependent) and subsequent `recycling' in a binary system
\citep{Alpar1982}.   The lowest mass NS are also potentially extremely
interesting.  If these systems are formed by electron capture supernovae \citep{Kitaura06,Ferdman14}, this could place a tight constraint on the EOS in terms of the relation between gravitational and baryonic mass \citep{Podsiadlowski05}.  This idea is, however, subject to uncertainties in both the
astrophysical setup (such as the progenitors), and on the nuclear physics
side with respect to the nuclear reactions during the SN and the EOS used for the
extraction of the masses.

SKA1-MID and SKA1-LOW will conduct both wide-field and targeted pulsar
surveys.  It is expected that an all-sky survey will discover $\sim 10000$
pulsars \citep{Keane14}, of which $\sim 1800$ will be MSPs and $\sim 100$ 
DNS.  Though impossible to predict precisely, $\sim$ 10\% of these should be well-suited for
EOS constraints.  SKA1 will also do deep targeted searches of
unidentified {\it Fermi} $\gamma$-ray sources \citep{Keane14} and
the Galactic globular cluster systems \citep{Hessels14}.  These
searches will discover primarily MSPs, some of which
will also be EOS laboratories.   The current spin record holder resides in the globular cluster Terzan 5 \citep{Hessels2006}.  A significant fraction of the
MSPs found in targeted searches will be in
eclipsing binaries (`black widows' and
  `redbacks' \citep{Roberts2011}), which, although more difficult to
detect and time precisely, may harbour the most massive
NS created by nature \citep{vanKerkwijk2011,Romani2012}
because they have accreted significant mass from their
companion and/or were formed more massive.

\section{State of the art (observations and theory)}
\label{sota}

\subsection{State of the art:  masses}
\label{sotamass}

\begin{figure}
\centering
\includegraphics[width=13cm]{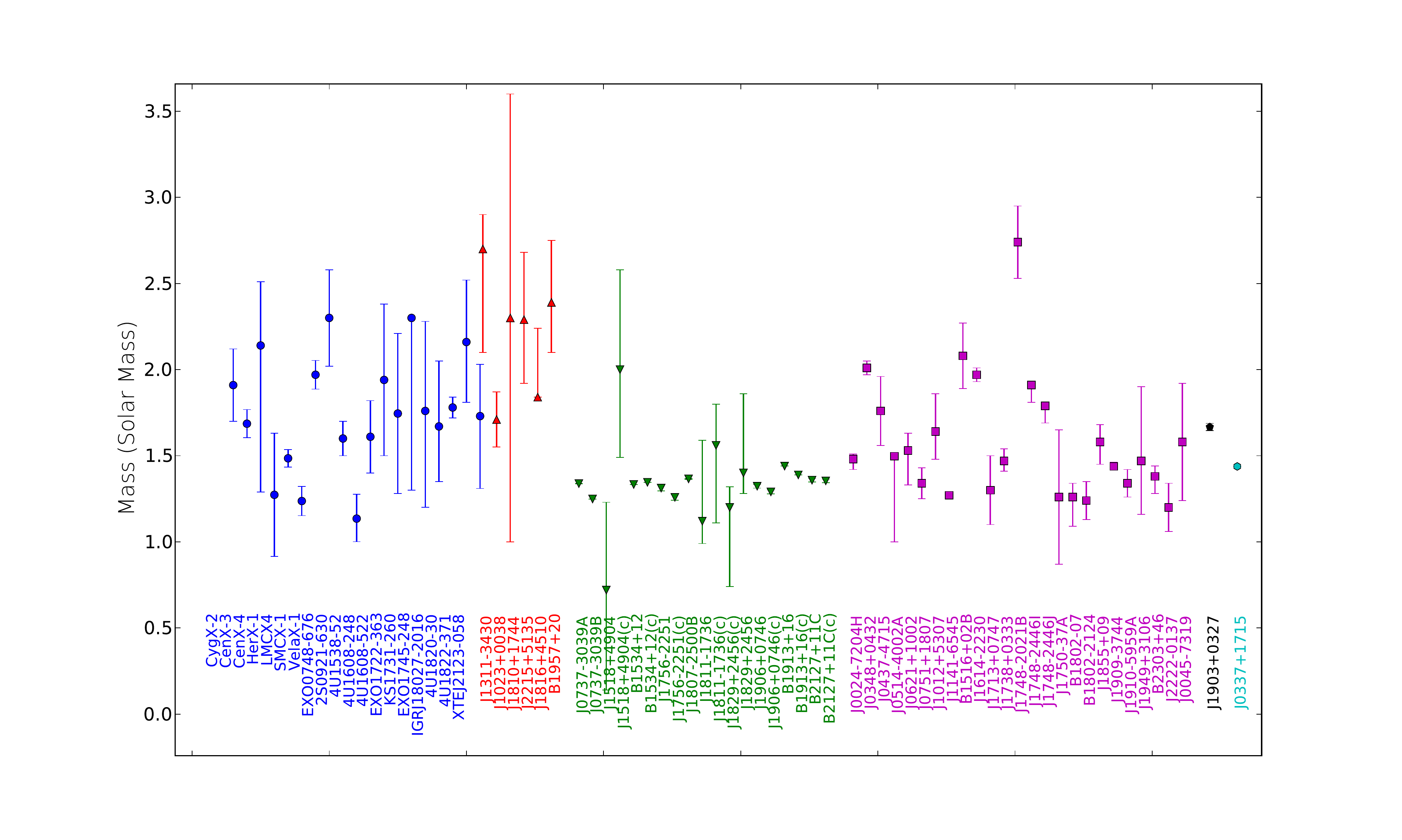}
\caption{Current NS mass measurements, 68\% confidence intervals (http://jantoniadis.wordpress.com/ research/ns-masses/ for original references). Blue -LMXBs; Red - Eclipsing MSPs; Green - DNS; Magenta - NS-WD binaries; Black - MSP-Main Sequence binaries; Cyan - Triple Pulsar. Though some sources may have higher masses than PSRs J1614-2230 and J0348+0432, there are large systematic uncertainties
  on these measurements or other reasons why they are not as
  direct or reliable a measurement.}
\label{nsmass}
\end{figure}

A measurement of a NS mass, even without a simultaneous measurement of $R$, constrains the EOS if it exceeds the maximum mass predicted by a given model (Figure \ref{eosmr}).  Figure \ref{nsmass} shows current NS mass measurements.  The two most constraining systems are PSRs\,J1614$-$2230 \citep{Demorest2010} and J0348+0432 \citep{Antoniadis2013},  with masses of $1.97\pm0.04$ and $2.01\pm0.03$\,M$_{\odot}$ respectively.  PSR\,J1614$-$2230 is a 3.1\,ms MSP orbiting an intermediate-mass WD every 8.7\,days. The system has a very high inclination ($89.17^\circ$) resulting in a strong Shapiro delay signal. PSR\,J0348+0432 is a 39\,ms pulsar in a relativistic, 2.46\,h orbit with a low-mass WD. For this  binary the PK parameters cannot be measured and the masses have been measured through phase-resolved optical spectroscopy of the bright WD companion.

Current knowledge of hadronic physics suggests it is nearly impossible
for matter to be purely nucleonic at the expected massive NS core densities of (7-10) $\rho_\mathrm{sat}$.  No matter what the precise composition, all alternative possibilities add additional degrees of freedom. The immediate effect is a softening of the EOS, lowering the maximum mass,
unless the interaction of the new constituents provides the necessary repulsion to counter-balance the strong gravitational
force. While 2$M_\odot$ NS do not directly exclude any components in the NS interior they do,
however, put very stringent constraints on the respective interactions.  In particular, they have triggered intensive discussions on the "hyperon
puzzle": ab-initio many-body calculations predict the appearance of hyperons
at 2-3 $\rho_\mathrm{sat}$ but the maximum masses of such models do not exceed  $\sim$ 1.7 $M_\odot$.  Proposed solutions include
a very early transition to quark matter, additional repulsion in the ab-initio
calculations via two- or more-body hyperonic forces and the appearance of
other new particles such as $\Delta$-baryons.

Evidence for even more massive NS has been reported for eclipsing binary pulsars with low-mass companions \citep{vanKerkwijk2011,Romani2012}. Based on optical observations of the companions, these systems seem to host NS with $M\sim 2.4 M_\odot$, well above the predictions of many EOS models.  However, there are fairly large systematic uncertainties associated with lightcurve modeling (which mostly impact the determination of inclination), rendering these measurements so far unreliable for EOS constraints. A significant improvement of these measurements will most likely require synergy of sensitive radio and optical observations with next-generation observatories such as SKA and E-ELT \citep{Antoniadis14}.

\subsection{State of the art:  moments of inertia}
\label{sotainertia}

Moment of inertia $I$ is a function of $M$ and $R$, so yields $R$ if $M$ is known independently. Current constraints on $I$ are very
approximate, and are based mostly on $\gamma$-ray flux measurements.
PSR~J1614$-$2230 is an especially interesting case,
since its mass is well known. This pulsar was among the 
first MSPs detected in $\gamma$-rays \citep{aaa+09}.   Assume that the spin down luminosity $\dot{E}  = 4 \pi^2 I \dot{P}/P^3$ \citep{Lorimer2005}.  Since the measured $\gamma$-ray luminosity must be $<$ 100\% of $\dot{E}$, this yields a lower limit on $I$.  If the distance to PSR~J1614$-$2230 is given by the free-electron model of the Galaxy \citep{cl02}, this yields $I > 10^{45}\, \rm g \, cm^{2}$ (not very constraining).   Distance estimates, which fix the $\gamma$-ray luminosity, are however uncertain.  The distance is also necessary to evaluate
 kinematic contributions (which may be large) to $\dot{P}$.  Detailed knowledge and modelling of the emission geometry is also necessary to
estimate the real luminosity of the pulsar: the estimate above assumes isotropic emission, which is not appropriate since in this case there would be no pulsations. One advantage of this method, however, is that such lower limits
on $I$ will eventually be available for a variety of other pulsars
\citep{gfc+12} with a variety of masses, something that might improve its stringency.

\subsection{State of the art: spin rates}
\label{sotaspin}

\begin{figure}
\includegraphics[width=15cm]{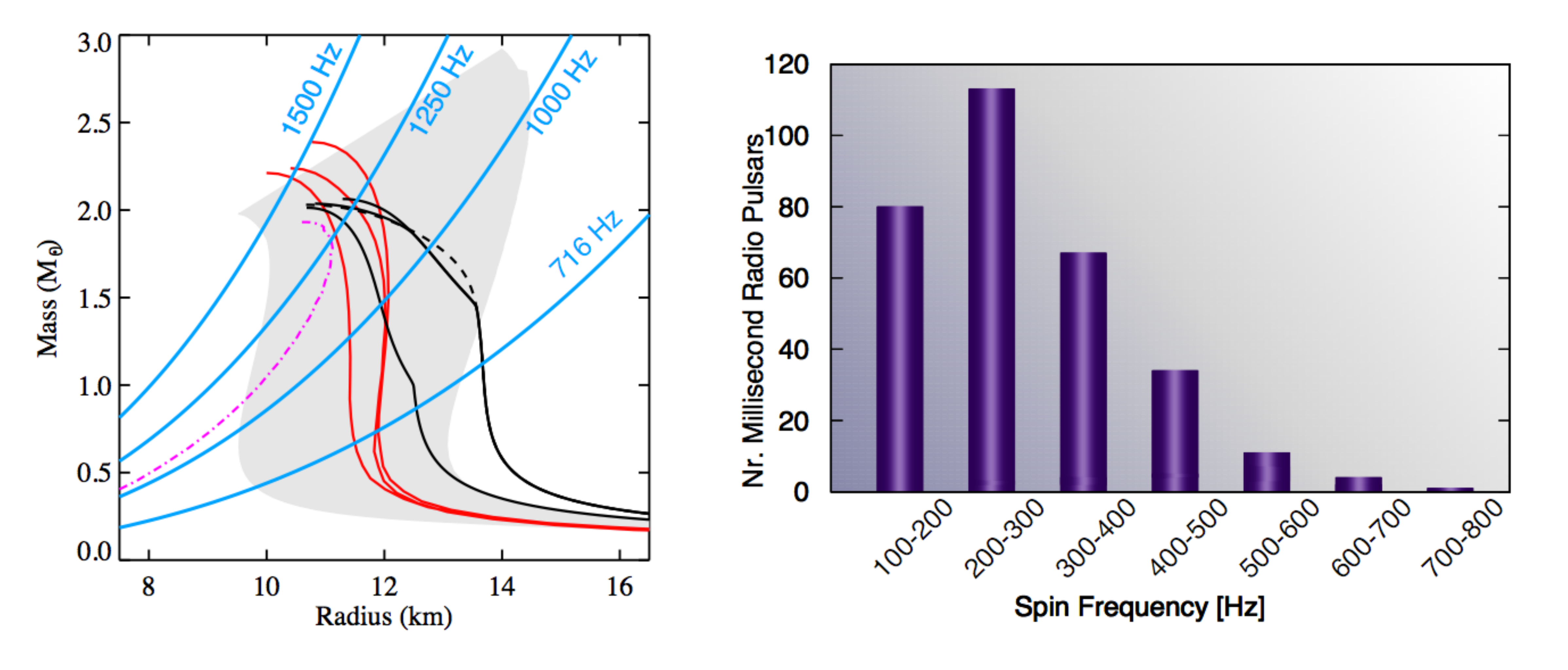}
\caption{Left:  Constraints arising from different spin rate measurements. NS of a given spin rate must
lie to the left of the relevant limiting line in the M-R plane (shown in blue for various spins). EOS models as in Figure 3.  The current record holder, which spins at 716 Hz \citep{Hessels2006} is not constraining, but given a high
enough spin individual EOS can be ruled out. Between 1 kHz and 1.25 kHz, for example, individual EOS in the grey band of nucleonic EOS would be excluded.  Right: The spin distribution of radio MSPs with a spin frequency $>$ 100 Hz.  The spin distribution of the smaller sample of accreting NS is similar \citep{Patruno2012,Watts2012}. }
\label{spincons}
\end{figure}

During recycling, a bare non-magnetic NS could easily attain very rapid spin after accreting only $\sim 0.1 M_\odot$ \citep{cst94,bpcdd99}\footnote{Although there are limiting mechanisms that might halt spin up, see \cite{Patruno2012} for a review.} .  Irrespective of the proposed EOS, the spin frequency $f$ must be lower than the Keplerian frequency $f_K$ above which a NS, due to centrifugal forces, becomes unstable to mass shedding at its equator.    To a very good approximation
$f_K = C (M/M_\odot)^{1/2} (R/10 \mathrm{km})^{-3/2}$ 
\citep{hzbl09}. where $M$ is the (gravitational) mass of the Keplerian configuration and $R$ denotes the (circumferential) radius of the non-rotating configuration of the same gravitational mass. $C$ is 1.08 kHz for NS and 1.12 kHz for strange stars~\citep{hzbl09}, obtained from fits to general relativistic calculations of rotating stars.   Since we require $f < f_K$, this results in a constraint on the radius: 

\begin{equation}
R < 10 ~C^{2/3} \left[\frac{M}{M_\odot}\right]^{1/3} \left[\frac{f}{1~\mathrm{kHz}}\right]^{-2/3}  ~\mathrm{km}
\end{equation}
The more compact the star is, the higher the supported rotation rate can be, and the tighter the constraint on the EOS.  Figure \ref{spincons} shows the spin distribution of the known radio pulsars.  The binary MSP J1748$-$2446ad \citep{Hessels2006} is the NS with the shortest known rotational period, $P_{min}=1.396$ ms, corresponding to a spin frequency of 716 Hz (the most rapidly rotating accretion-powered NS rotates at 619 Hz \citep{Hartman03}).   The resulting constraint is shown in Figure \ref{spincons}:  the radius of a non-rotating 1.4 $M_\odot$ star would have to be smaller than 15 km.   This is not particularly constraining for current models:  rotation rates faster than a millisecond are required to rule out EOS.  

\subsection{State of the art:  glitches}
\label{sotaglitch}

\begin{figure}
\centering
\includegraphics[width=10cm]{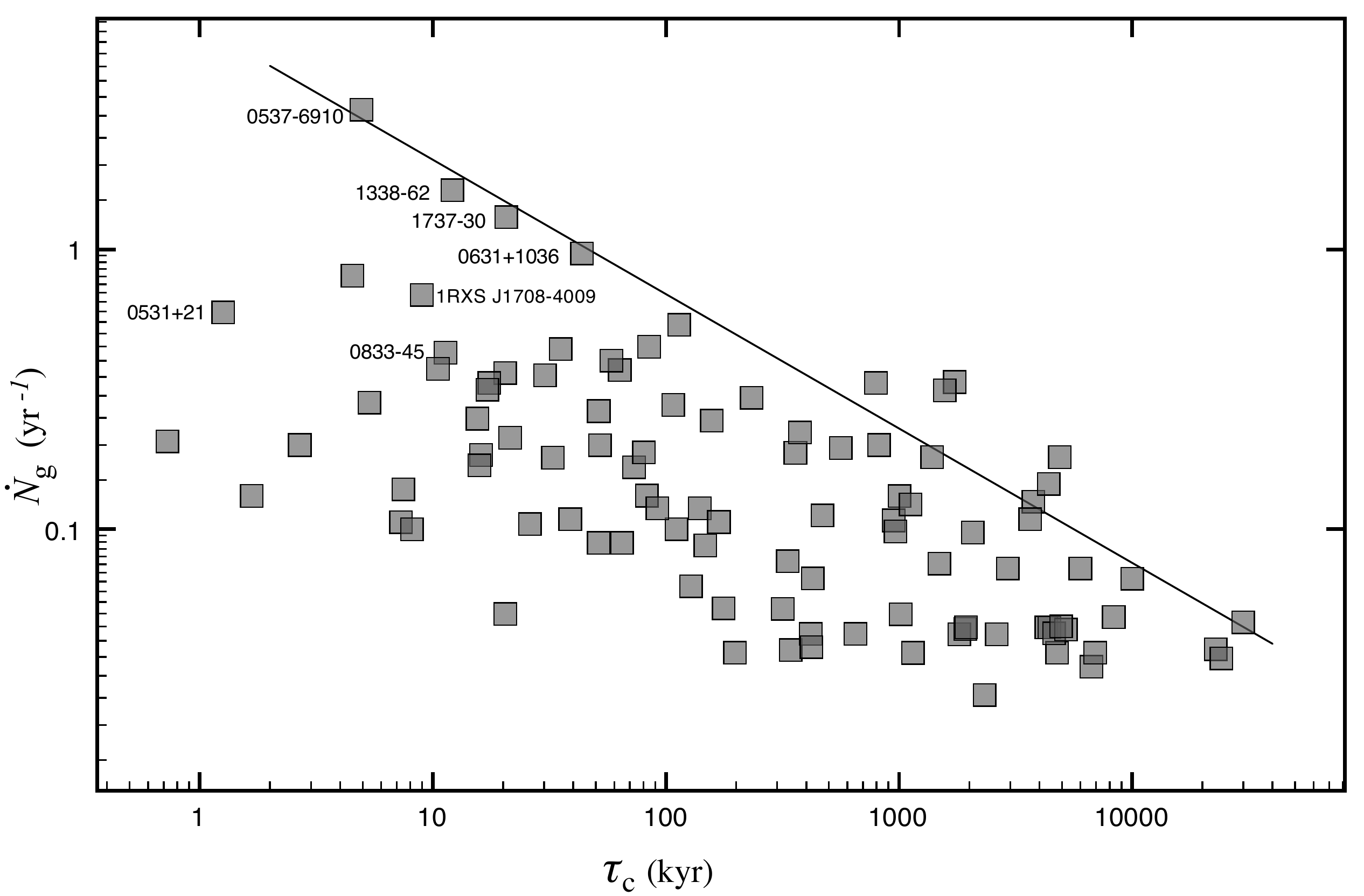}
\caption{Number of glitches per year for most pulsars known to have glitched and observed for more than 3 yr (by 2010, from \cite{elsk11}).  The straight line is a linear fit to the maximum value of $\dot{N}_g$ in each half-decade of characteristic age $\tau_c$.}
\label{longpulse}
\end{figure}

The majority of the glitches known today have been detected in the data of long monitoring programs (at Jodrell Bank, Parkes and Urumqi; \citep{ywml10,elsk11,ymh+13}). This is because glitches are inferred from observations before and after the glitch epoch.  The most minimal characterisation of a glitch is the measurement of the step in spin frequency ($\Delta\nu$).  However, many glitches also induce a change in spin-down rate ($\Delta\dot{\nu}$) and a process of relaxation towards the pre-glitch rotational values, with timescales that can go up to a few hundred days. 
Constraints and information on the glitch mechanism and the rotational dynamics of the neutron superfluid can be obtained if we can detect and characterise many (if not all) glitches in several sources, precisely.  Our ability to do this depends directly on the cadence of the observations, as well as on the accuracy of the TOAs.     At the moment we know of $\sim$ 440 glitches in $\sim$150 pulsars. Glitch activity is concentrated in young objects ($\tau_c<100$\,kyr) and there is an almost linear correlation with spin-down rate, i.e. the faster a pulsar spins down, the higher its glitch activity \citep{lsg00, elsk11}. The occurrence of glitches is in general not periodic and most pulsars exhibit less than one glitch every 2 years.  This shows the importance of monitoring a large number of pulsars over a long time (Figure \ref{longpulse}) using sub-arraying to observe multiple sources at higher cadence when possible.

The size of the largest glitches has already constrained the amount and location of the participating superfluid \citep{aghe12,cha13,pfh14}.  The detection of small glitches and the lower end of the $\Delta\nu$ distribution are however highly uncertain. Below certain sizes detection is compromised and there is confusion with phenomena like timing noise and disperson measure (DM) variations.  A study of the glitch activity of the Crab pulsar has shown that the number of glitches decreases significantly below $\Delta\nu=0.05\,\mu$Hz, implying a rather large minimum glitch size \citep{eas+14} which can in principle constrain the trigger mechanism.  However, the minimum size found for the Crab glitches is larger than many glitches detected in other pulsars.  In-depth studies of other pulsars will be important to assess the universality of the existence of a lower cut-off and to reach a better comprehension of the glitch trigger mechanism.
To accomplish this it is necessary to have observations sensitive to very small glitches: the Crab pulsar glitch study was enabled by the high cadence (one observation per day) and high precision of the data.

Glitch recoveries provide information on the dynamics of the internal superfluid and its interaction with the other NS constituents. The Vela pulsar is the prototype source showing glitch recoveries, which are also observed in many other young pulsars. During the relaxation process the spin-down rate slowly decays from the rather large (very negative) value induced by the glitch. It takes years for it to again reach pre-glitch levels and in many cases the process will be interrupted by a new glitch. PSR B2334+61 exhibited one of the largest glitches known to date, followed by a standard recovery process.  Nonetheless, significant oscillations were reported in post-glitch timing residuals \citep{ymw+10}. To date no post-glitch oscillations have been detected in other pulsars.  This could be related to the size of the glitch but could also just be an observational bias against detecting similar effects in lower sensitivity data taken at lower observing cadence..

Other pulsars have shown peculiar glitch recoveries, like the RRAT PSR J1819$-$1458 and the radio pulsar PSR J1119$-$6127 \citep{lmk+09,wje11}. In these two cases the spin-down rate over relaxed and reached values significantly smaller than the pre-glitch levels. Pulse shape variations were detected in a couple of observations following the glitch in PSR J1119$-$6127 but the rather low cadence available (month$^{-1}$) prevented a more detailed study. The unique recoveries were attributed to the large dipole fields of the two pulsars, though no physical mechanism was proposed.  The analysis of glitch recoveries can tell us about differences between the regions responsible for the rotational relaxation, for instance whether there is pinning in the outer core or not. These studies would benefit from higher cadence and more precise observations.

\section{Expectations from SKA}
\label{ska}

\subsection{Improvements in mass measurements}
\label{impmass}

\begin{figure}
\centering
\includegraphics[width=10cm]{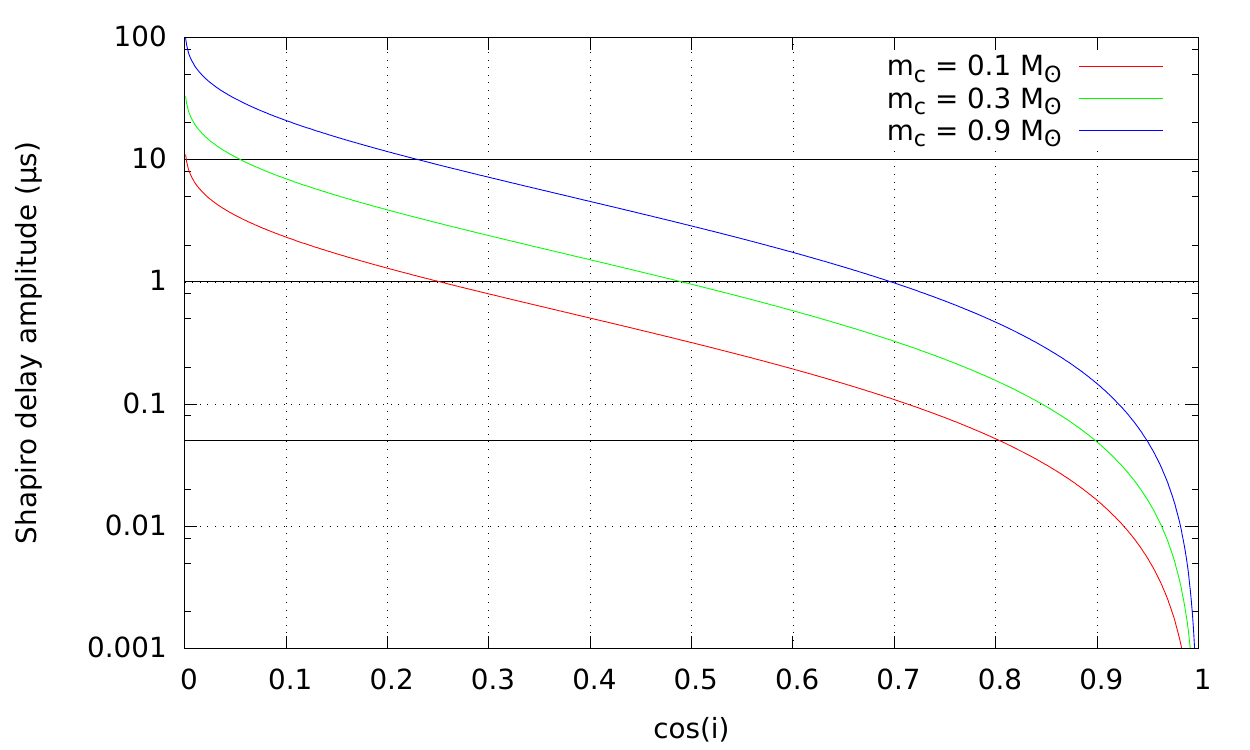}
\caption{
        Observable Shapiro delay amplitude -- i.e., the portion of the
        signal that is not covariant with other orbital parameters -- as
        a function of inclination $i$ and binary companion mass.
        Assuming orientations have uniform $\cos i$, at timing
        precision of $\sim$50~ns, $\gtrsim$80\% of NS-WD binary systems
        with $m_c \ge 0.1$~M$_\odot$ should show detectable Shapiro
        delay.
}
\label{fig:shapiro_amp}
\end{figure}
 
The SKA is expected to improve on the number of measured NS
masses in two ways.  First, the dramatic increase in the
number of known radio pulsars will yield more exotic systems in which masses can be determined (e.g. via measurable Shapiro delay).  We might expect these to comprise roughly the same fraction
of the total population as is observed now, or $\sim$1--2\%.  Even so,
this effect alone would increase the number of total mass measurements
by a factor $\sim$ 10.  The second way that the SKA will
improve mass measurements is through the improved timing precision that
will be achievable with the massive increase in telescope sensitivity.
This will improve the uncertainty on already-measured masses.  It also
has potential to increase the {\it fraction} of systems in which masses
can be measured, boosting the total numbers even further.  Detection of Shapiro delay in significant fractions of the NS-WD population may become possible (Figure \ref{fig:shapiro_amp}).

In contrast to many of the other relativistic timing parameters, Shapiro
delay is not heavily dependent on special environmental or evolutionary
history.  The effect applies to all binary systems, is insensitive to
contamination by non-relativistic effects, and its detectability mainly
depends on random geometrical orientation -- at high inclinations, the
radio pulses we receive must travel close to the companion star,
resulting in a much larger time delay\footnote{Note however that detectability of a Shapiro delay would be adversely affected if the inclination were so high that the system was in fact eclipsing.}.  Furthermore at low
inclinations, the effect becomes highly covariant with the Keplerian
orbital parameters, which reduces its observable amplitude even further \citep{Freire2010}.  Nevertheless, if high enough timing precision can be achieved, the effect can be detected even in fairly
low-inclination systems.  This is illustrated in
Figure~\ref{fig:shapiro_amp}, which shows the observable (non-covariant)
Shapiro delay amplitude versus $\cos i$.  At $\sim$50~ns timing
precision, Shapiro delay is potentially detectable at inclinations as
low as $\sim 40^\circ$ -- this includes $\sim$80\% of all binary
systems assuming random orientations (uniformly distributed in $\cos
i$).   The SKA is expected to achieve sufficient timing precision, and discover enough new binary
systems, that we might expect at least a factor of five more Shapiro delay based mass measurements. 

\subsection{Improvements in moment of inertia measurements}
\label{impinertia}

The SKA opens up the possibility of a precise measurement of
moment of inertia $I$ for one pulsar for which the mass is already
precisely known, the Double Pulsar PSR~J0737$-$3039A.  There are three routes by which pulsar timing measurements can lead to a measurement of  $I$ \citep{kw09}.  A highly relativistic system such as the Double Pulsar \citep{bdp+03,lbk+04} is required in order for the effects of $I$ on the pulsar timing to be discernable.  The first measurable effect is expected to be a contribution comparable in size to the 2PN-level correction to the advance of periastron, $\dot \omega$.  This would produce an offset between the measured value of $\dot \omega$ and that expected by taking the other measured PK timing parameters and combining them for a (1PN) prediction.  In practice, this will require very precise measurement of at least two other PK parameters which produce largely orthogonal curves in the mass-mass plane.  For the Double Pulsar, the parameters best suited to this prediction will be the sine of the orbital inclination angle, $s$, and the rate of decay of the orbital period, $\dot P_{\rm b}$.  Simulations assuming 5\,$\mu$s timing precision predict a 10\% measurement of $I$ in 20 years \citep{kw09}.  Timing with SKA1 should do even better than this.  An important point is that the corrections to $\dot P_{\rm b}$ for the relative acceleration of the Solar System and the pulsar system must be known to high accuracy in order for this method to work.  These corrections require precise measurements of the pulsar's proper motion and distance, things that should be easily achieved for the Double Pulsar with the timing and VLBI capabalities of SKA1.  A further issue is the need for a reliable model of the Galactic potential; as the Double Pulsar is close to the Plane, this should not pose an insurmountable problem in this particular case, and models will also be improved on the necessary timescale by the results of the GAIA satellite.

The second method involves determining non-linear time evolution of the longitude of periastron, if the spin of the (recycled) pulsar is inclined with respect to the system's angular momentum.  This should allow the measurement of the projection of the pulsar's spin vector on the orbital plane, leading to $I$ via a precession-determined estimate of the spin-orbit misalignment angle $\delta$.    The third technique relies on measuring two time derivatives of the projected semi-major axis of the pulsar's orbit, $x$, once again leading to an estimate of $I\sin \delta$, although this effect would need to be separated from aberration-induced changes in projection of the orbit \citep{ds88,wk99,dt92}.  For the Double Pulsar these last two techniques are not promising, as the spin-orbit misalignment angle for the recycled pulsar is known to be small \citep{fsk+13} and furthermore the orbit happens to be nearly edge-on to the line of sight, making the changes in projection tiny.  The small misalignment angle is likely the result of a fairly symmetric, low-mass-loss explosion in the second supernova in the system \citep{plp+04}, but there are other DNS systems that are known to be significantly misaligned \citep{kra98,wt02,fst14}.  The SKA (1,2) is expected to discover (100,180) DNS systems, and it is reasonable to hope that favourable geometries, and indeed even more relativistic orbits will be found in some of them.

 \subsection{Improvements to spin measurements}
\label{impspin}

In the more than three decades since the discovery of the first MSP B1937+21 ($P=1.56$ ms) \citep{bkhdg82},  only one object, PSR J1748$-$2446ad ($P=1.39$ ms) \citep{Hessels2006} has been found that spins faster. Until a few years ago, the sometimes prohibitive computational costs of  a large survey sensitive to very rapidly rotating objects lent support to the hypothesis that there might have been an observational bias limiting the number of known MSPs with spin period close to that of the record holders or below. However, the improvement of the computational capabilities implemented in modern pulsar search experiments made them nominally able to pick up ultra-fast radio pulsations and several new binary MSPs with spin periods below 2 ms have been discovered in recent years. An additional contribution to the increment of this population of rapidly spinning NS came from the targeted radio searches towards $\gamma$-ray unidentified sources selected from the {\it Fermi} catalogues.

Interestingly, in 9 cases (over a total of 11 binary MSPs having $P<2$ ms), including the spin record-holder PSR J1748$-$2446ad, the most rapidly rotating pulsars appear to be hosted in eclipsing systems: binaries in which the matter from the companion obscures the radio pulsations for a large fraction of or perhaps even the entire orbit.   This suggests that although there may be no observational bias against discovering the most rapidly rotating systems, they be nonetheless be hidden by obscuration from the companion star. However, the eclipses in these systems appear to be strongly frequency dependent, with the eclipses being significantly longer at frequencies at a GHz and below.  It is also interesting to note that many pulsars in eclipsing systems were discovered in a 2 GHz survey of globular clusters with the GBT. 

Due to the very large numbers of beams needed to cover the sky and accounting for the long dwell time needed to reach the required sensitivity (a pulsar like J1748$-$2446ad has a flux density of only $\sim 80$ $\mu$Jy at 2 GHz) a deep all-sky survey at 2 GHz is not feasible with the current generation of radio telescopes. The beam-forming capabilities (allowing simultaneous search of tens or hundreds of sky positions), and the higher instantaneous sensitivity of SKA1-MID  (reaching $\sim 10$ $\mu$Jy sensitivity in a relatively short integration time) will give the first opportunity for a deep all-sky survey at 2 GHz\footnote{If observing time does not permit a full survey, one would instead focus on pointed observations of GCs, {\it Fermi} sources, and quiescent LMXBs to find the fastest spinning pulsars.}. Provided the sub-millisecond pulsars exist and are mostly hidden in eclipsing binaries, this experiment would be the best designed ever in the radio band with the aim of uncovering at least the brightest objects in this category. A similar later experiment performed with SKA2 will be finally sensitive even to sub-millisecond pulsars in a portion of the Galaxy well beyond the volume probed by the {\it Fermi} $\gamma$-ray satellite.  

Even if SKA1-MID and SKA2 do not find sub-millisecond pulsars, SKA1-MID will certainly discover at least few MSPs spinning faster than PSR J1748$-$2446ad, while SKA2 will deliver a full census of the population of the of the nearby, thus unaffected by scattering and dispersion delays, ultra rapidly spinning MSPs. Likely, the accurate timing observations possible with SKA will also lead to a measurement of the pulsar mass (via relativistic effects) for at least a sub-group of the objects. It will then be possible to exploit the large sample of rapidly spinning sources in order to derive statistically sound results. Population analysis \citep{pcg+99,c08,hma11,ptrt14} may constrain not only the EOS, but will also elucidate other fundamental physical processes related to the final stage of the recycling that may depend on the EOS e.g.: {\it (i)} the role of  the emission of gravitational wave radiation (r-modes or accretion induced quadrupole moments  \citep{b98,bw13})  in halting the pulsar spin up; {\it (ii)} the physical process provoking the decay of the surface magnetic field in the MSPs; {\it (iii)} the relative importance of the radio ejection \citep{bpdd+01} and of the Roche-lobe decoupling phase \citep{tau12} in shaping the spin distribution of the MSPs (see also \cite{kt10}).

The impact of a NS rotating at over $\sim$ 1.1 kHz would be
substantial, and 1.5 kHz would be revolutionary: no current models can sustain such a fast spin. A hypothetically observed frequency of 1.4 kHz, for example, would constrain the radius of a 1.4 $M_\odot$ NS  to less than 9.5 km (Figure \ref{spincons}).  In general higher frequencies favour additional non-nucleonic degrees of freedom, since they give more compact stars.  Simultaneous measurement of high mass and high spin would be particularly constraining \citep{hzb08}.  While massive NS
require repulsion among their constituents at high density, fast rotation requires the interaction not to be too repulsive to counter-balance
the centrifugal force. 

\subsection{Improvements in glitch sampling}
\label{impglitch}

The detection of glitches and the characterisation of their recoveries rely strongly on the constant monitoring of a sample of pulsars.  The observing capabilities which are currently offered by the radio observatories that monitor young pulsars are sufficient for the detection of medium to large glitches.  
The detection of small glitches and the study of the lower end of the glitch size distribution are, however hard to attain with the current observing programs.
SKA glitch studies should focus on monitoring a selected group of pulsars and should be able to provide the complete glitch sample for the observed sources during the observing campaign. Besides uncovering the lower end of the size distribution, this will allow the study of waiting time (inter-glitch time intervals) distributions, which can also shed light on the glitch mechanism. 

There are two main factors limiting the detection of glitches \citep{eas+14}:
\emph{1)} How often pulsars are observed, and
\emph{2)} $\sigma_\phi$, the larger of the dispersion of the timing residuals in a given time-interval (10-15 observations) and the average uncertainty of the TOAs.
Compared to the current observing setups the SKA will offer superior sensitivity, hence we assume that $\sigma_\phi$ will never be dominated by TOA uncertainties.  Unmodelled rotational irregularities and observing cadence will likely be the main limiting factors in glitch detection.  Among the known effects that can increase the dispersion of the timing residuals are DM variations and timing noise.  DM variations will be corrected via regular multi-frequency observations (highlighting the need for SKA1-LOW), necessary to achieve high precision timing, and therefore should not affect glitch detection.  Timing noise, on the other hand, still needs to be well understood in order to correct the data for its effects.  The detection of small glitches (for a given pulsar) will therefore be dependent on the intrinsic levels of timing noise and its characteristic timescales in combination with the observing cadence.

The nominal cadence of SKA observations for new pulsars is 1 observation per 2 weeks, for at least a year.
In addition, timing campaigns for pulsars should last at least 10 years.  Although this might seems appropriate, with one observation every two weeks the SKA might not be able to improve substantially the statistics of small glitches already accomplished by some of the current programs, nor to sample satisfactorily post-glitch recoveries. Glitch studies need a higher cadence and a longer campaign over a fixed set of pulsars.  More frequent SKA observations are planned for a sub-sample of interesting MSPs, but these are expected to glitch at an average rate lower than 1 glitch per millennium \citep{elsk11}.   More complete monitoring of a larger sample of pulsars may be achievable using SKA1-LOW, which, with the wider field of view and sufficiently large numbers of tied-array beams, should facilitate daily observations of large numbers of sources. 

\begin{figure}
\begin{center}
\includegraphics[height=7cm]{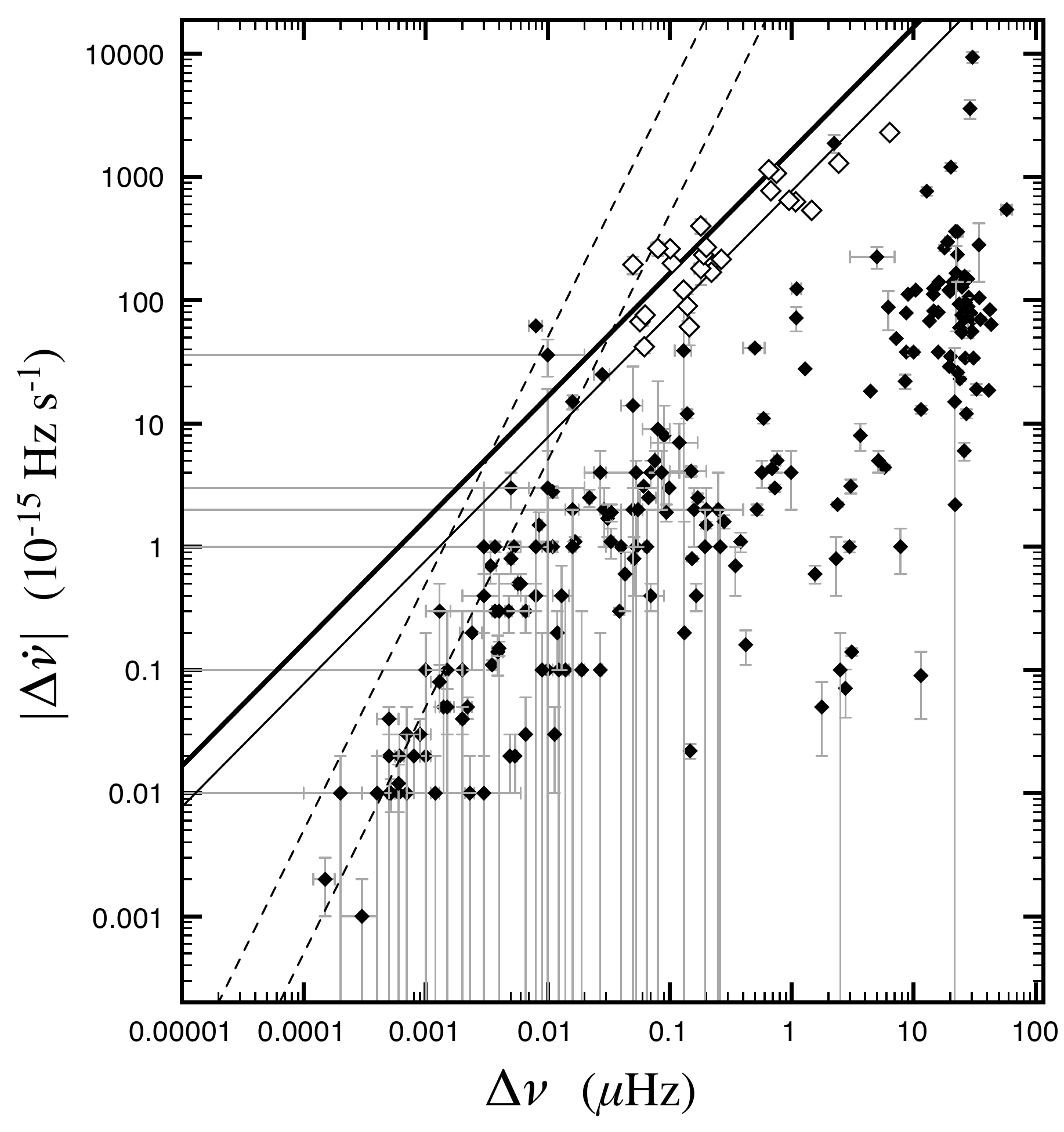}
\includegraphics[height=7cm]{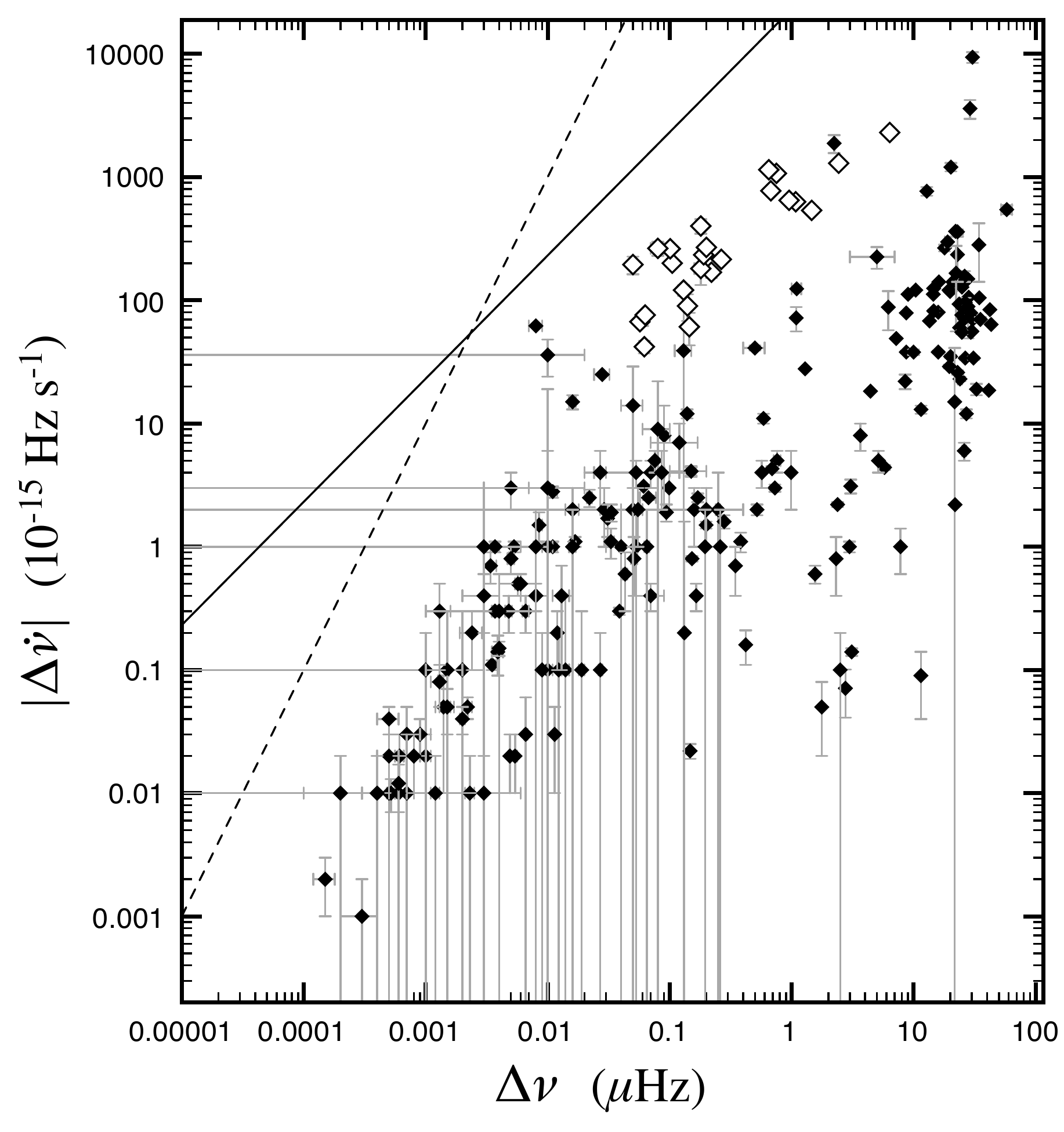}
\end{center}
\caption{245 glitches in the $|\Delta\dot{\nu}|$--$\Delta\nu$ plane \citep{elsk11}. Open diamonds: Crab pulsar glitches.
\emph{Left}: Detectability limits for observing cadences of 14 days (solid thick line)  and 30 days (solid line). The segmented lines are for $\sigma_\phi$ equal to $0.01$ (rightmost) and $0.001$ (leftmost) rotations, which are values commonly reached today.
\emph{Right}: Same as \emph{Left} but for a cadence of 1 day and TOA precision of $5\times 10^{-5}$ rotations. 
}
\label{glits}
\end{figure}

Glitch detectability limits also depend on the size of the spin-down rate steps, improving for smaller steps.  Known glitches have $|\Delta\dot{\nu}|$ values in the range $10^{-3}$ to $10^{4}$ ($\times10^{-15}$\,Hz\,s$^{-1}$).  The two-weeks cadence strongly limits the detection of glitches with  $|\Delta\dot{\nu}|>1\times10^{-15}$\,Hz\,s$^{-1}$.  For instance, some of the Crab pulsar glitches could not be unambiguously detected, as demonstrated in Figure \ref{glits}. The bias in glitch detection is clear from the left hand side plot, being dominated by $\sigma_\phi$ at low $|\Delta\dot{\nu}|$ values.
For some pulsars $\sigma_\phi$ is driven by timing noise, but for others is driven by TOA precision.
The right hand side plot shows the increased parameter space that SKA observations could achieve with daily observations and a timing precision of at least $5\times 10^{-5}$ rotations (equivalent to $5\times10^3$\,ns, for a period of $0.1$\,s). The SKA could easily achieve this precision for many pulsars, but the level of timing noise present in a given time-interval (10-15 TOAs) could be higher. 

The number of pulsars monitored for glitches need not be large (20--30 objects) and should include those that hav
e already been monitored for years (to track long-term glitch history), which are bright for SKA standards. Precision can thus be traded for higher cadence. The Aperture Array systems should allow an optimum combination of sensitivity, field-of-view and number of beams to yield exceptional cadence on a very large number of sources.  With one \emph{short} ($\sim$ 2 mins\footnote{Assuming $A_{eff}/T_{sys}=1000$\,m$^2$\,K$^{-1}$ (SKA 1) and for a duty cycle of $10$\%, average flux $0.5$\,mJy and a bandwidth of $500$\,MHz, $2$ mins. of observation gives $2\times10^{-5}$\,s precision (for P=$0.1$\,s)}) observation per day giving a TOA precision of at least $5\times 10^{-5}$ rotations, over years and for a sample of pulsars, we can explore fully the glitch size distribution and track the recoveries of all detected glitches.  SKA will thus yield the ideal sample to probe the glitch mechanism.

\section{Summary}
\label{implications}

The nature of the strong interaction is fundamental to understanding the behaviour of systems from nuclei to neutron stars, and remains a serious challenge to physicists even though it is mathematically well-defined.    SKA1, and ultimately the full SKA, will progressively give us the opportunity to explore the nature of matter, the strong force and superfluidity under the extreme conditions that prevail inside NS, via precise measurement of pulsar masses and moments of inertia, the discovery of rapidly-spinning and new types of pulsars, and high cadence observations to detect pulsar glitches.


\end{document}